\documentclass{jfm}
\usepackage[T1]{fontenc}
\usepackage[latin9]{inputenc}
\usepackage{soul}
\usepackage{xcolor}
\usepackage{graphicx}
\usepackage{amssymb}
\usepackage[authoryear]{natbib}

\makeatletter

\providecolor{lyxadded}{rgb}{0,0,1}
\providecolor{lyxdeleted}{rgb}{1,0,0}

\ifCUPmtlplainloaded \else
  \IfFileExists{amsbsy.sty}
    {\typeout{^^JFound the 'amsbsy' package on the system, using it.^^J}
}

\def\vec#1{\mbox{\boldmath ${#1}$}}
\def\Bm{\textit{Bm}}
\def\d{\mathrm d}
\def\e{\mathrm e}
\def\i{\mathrm i}
\def\citet#1{\cite{#1}}

\affiliation{Applied Mathematics Research Centre, Coventry University, Priory Street, Coventry CV1 5FB, UK}
\pubyear{2011}
\volume{0}

\begin{document}

\title[Edge pinch instability of oblate liquid metal drops]
{Edge pinch instability of oblate liquid metal drops in a transverse AC magnetic field}

\author[J. Priede, ]{J\ls \=A\ls N\ls I\ls S\ns P\ls R\ls I\ls E\ls D\ls E}

\date{20 April 2010; revised 20 December 2010; accepted 20 January
2011}

\maketitle

\begin{abstract}
This paper considers the stability of liquid metal drops subject to
a high-frequency AC magnetic field. An energy variation principle
is derived in terms of the surface integral of the scalar magnetic
potential. This principle is applied to a thin perfectly conducting
liquid disk, which is used to model the drops constrained in a horizontal
gap between two parallel insulating plates. Firstly, the stability
of a circular disk is analysed with respect to small-amplitude harmonic
edge perturbations. Analytical solution shows that the edge deformations
with the azimuthal wavenumbers $m=2,3,4,\ldots$ start to develop
as the magnetic Bond number exceeds the critical threshold $\Bm_{c}=3\pi(m+1)/2.$
The most unstable is $m=2$ mode, which corresponds to an elliptical
deformation. Secondly, strongly deformed equilibrium shapes are modelled
numerically by minimising the associated energy in combination with
the solution of a surface integral equation for the scalar magnetic
potential on an unstructured triangular mesh. The edge instability
is found to result in the equilibrium shapes of either two- or three-fold
rotational symmetry depending on the magnetic field strength and the
initial perturbation. The shapes of higher rotational symmetries are
unstable and fall back to one of these two basic states. The developed
method is both efficient and accurate enough for modelling of strongly
deformed drop shapes.
\end{abstract}

\section{Introduction}

In several metallurgical processes such as, for example, the levitation
melting and cold-crucible, where the induction heating is used, the
surface of liquid metal is subject to AC magnetic field. In such a
way, the metal can be not only molten but also evaporated provided
that the heating power is high enough (\citealt{BLGP07}). Induction
heating is accompanied with a pinch effect, which can significantly
deform the surface of liquid metal. When a sufficiently strong magnetic
field is applied, surface sometimes becomes asymmetric and even strongly
irregular (\citealt{FSE07}). This phenomenon is of primary importance
for the induction heating of liquid metals because it may have an
adverse effect on the heating efficiency and eventually limit the
power density the liquid metal can dissipate. Such a surface instability
has been observed first by \citet{PFE03} on a circular layer of Gallium
in a mid-frequency AC magnetic field. Analogous instability was studied
also by \citet{Moh05} on the free surface of InGaSn melt in the annulus
placed under a ring-like circular coil and fed by an alternating current
with the frequency in the range of $20-50\,\mbox{kHz}.$ As the current
amplitude exceeds a certain critical value, which depends mainly on
the annulus width, an initially flat surface acquires a static wavy
deformation. At a higher critical current, the deformation rapidly
increases and becomes unsteady. In contrast to \citet{PFE03}, who
observe only static deformations, \citet{KKCS06} find a circular
sessile drop of InGaSn melt first to squeeze radially with various
shape oscillations to set in as the strength of a $20\,\mbox{kHz}$
AC magnetic field is gradually increased. In the former experiment,
the surface of liquid metal was exposed to the air and, thus, heavily
oxidised that constrained its motion. In the latter experiment, oxidation
was prevented by covering the drop by a diluted HCl solution. Later
on \citet{CKK06,Con07} found static shape deformations when the drop
was constrained in a horizontal gap between two parallel plates. Irregular
static surface shapes have been observed also by \citet{HVD06a} on
a layer of PbSn alloy covering the bottom of a cylindrical container.
The metal layer, which was constrained by the lateral walls of the
container and heavily oxidised at the top, broke up revealing the
bottom of the container as the strength of a $4\,\mbox{Khz}$ AC magnetic
field exceeded a certain critical value. The authors also attempted
to model this process numerically using a surface integral equation
derived from Green's third identity. This approach, however, is not
applicable to thin sheets, for which the double layer contribution
vanishes. In a subsequent paper, \citet{HVD06b} devised a simplified
electrotechnical model, which provided a rough estimate of equilibrium
shapes. A simple theoretical model for this type of instability was
introduced by \citet{PEF06}, who analysed the linear stability of
the edge of liquid metal layer, which was treated as a perfectly conducting
thin liquid sheet in a transverse AC magnetic field. This allowed
the authors to determine the wavenumber of the most dangerous perturbation
and the critical field strength at which the instability develops
in a reasonable agreement with the observations of \citet{Moh05}.

In this paper, an energy variation principle is derived for the equilibrium
shapes that develop from the edge pinch instability of flat liquid
metal drops, which are modelled as thin perfectly conducting liquid
sheets. Firstly, the stability of a circular disk is analysed with
respect to small-amplitude harmonic edge perturbations. Analytical
solution shows that the edge deformations with the azimuthal wavenumbers
$m=2,3,4,\ldots$ start to grow as the magnetic Bond number exceeds
the critical threshold $\Bm_{c}=3\pi(m+1)/2.$ The most unstable is
$m=2$ mode, which corresponds to an elliptical deformation. Secondly,
strongly deformed equilibrium shapes are modelled numerically by minimising
the associated energy. The electromagnetic problem is formulated in
terms of the surface integral equation for the scalar magnetic potential,
which is solved numerically on an unstructured triangular mesh covering
the surface of the drop. The edge instability is found to result in
the equilibrium shapes of either two- $(m=2)$ or three-fold $(m=3)$
rotational symmetry depending on the initial perturbation and the
magnetic field strength. Although the associated energy of $m=3$
shapes is higher than that of $m=2$ ones at the same magnetic field
strength, both shapes are separated by a positive energy barrier.
This, however, is not the case for equilibrium shapes of higher order
symmetries. Although these shapes can be obtained numerically, they
turn out to be unstable with respect to small amplitude perturbations
of two- or threefold rotational symmetries, which make them fall back
to one of the two basic states.

This paper is organised as follows. In $\S$\ref{sec:probform}, the
problem is formulated and the energy variation principle derived in
terms of the integral of the scalar magnetic potential over the drop
surface. This principle is applied in $\S$\ref{sec:Anal} to obtain
an analytical solution for the stability of a circular disk with respect
to small-amplitude harmonic edge perturbations. Specific mathematical
details of the solution are given in Appendix \ref{sec:apx1}. In
$\S$\ref{sec:Num}, numerical method is described and validated against
the previous analytical solution. Numerical results are presented
in $\S$\ref{sec:Num}. The paper is concluded with a summary and
discussion in $\S$\ref{sec:Sum}.

\section{\label{sec:probform}Formulation of problem}

\begin{figure}
\begin{centering}
\includegraphics[width=0.5\columnwidth]{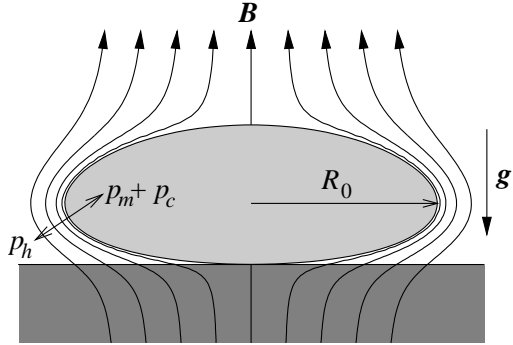} 
\par\end{centering}

\caption{\label{cap:sketch}Sketch to the formulation of problem.}

\end{figure}

Consider a drop of liquid metal with the characteristic size $R_{0},$
electrical conductivity $\sigma,$ surface tension $\gamma$ and density
$\rho$ submitted to an AC magnetic field with the spatial amplitude
distribution $\vec{B}(\vec{r}),$ , as shown in figure \ref{cap:sketch}.
The AC frequency $\omega$ is assumed so high that the penetration
depth of the magnetic field into the drop $\delta\sim(\mu_{0}\sigma\omega)^{-1/2}$,
where $\mu_{0}$ is the vacuum permeability, is negligible with respect
to $R_{0}.$ In this perfect conductor approximation, the magnetic
field is tangential to the drop surface $S,$ \begin{equation}
\left.B_{n}\right|_{S}=0,\label{eq:Bn}\end{equation}
 and the electromagnetic force effectively acts on the surface as
the time-averaged magnetic pressure, \[
p_{m}=\frac{\vec{B}^{2}}{4\mu_{0}}.\]
Equilibrium shape of the drop is determined by the normal stress balance
\begin{equation}
\left.p_{h}-p_{c}-p_{m}\right|_{S}=0,\label{eq:pblnc}\end{equation}
where $p_{h}=\rho\vec{g}\cdot\vec{r}$ and $p_{c}=\gamma\vec{\nabla}\cdot\vec{n}$
are the hydrostatic and capillary pressures, respectively, $\vec{n}$
is the outward surface normal and $\vec{g}$ is the gravitational
acceleration. Multiplying (\ref{eq:pblnc}) by $\vec{n}\cdot\vec{\xi},$
where $\vec{\xi}(\vec{r})$ is a virtual displacement field conserving
the volume, and integrating over $S,$ we obtain\begin{equation}
W_{g}+W_{s}+W_{m}=0,\label{eq:wblnc}\end{equation}
where $W_{g}=\rho\int_{s}(\vec{g}\cdot\vec{r})\vec{\xi}\cdot\d\vec{s},$
$W_{s}=-\gamma\int_{s}(\vec{\nabla}\cdot\vec{n})\vec{\xi}\cdot\d\vec{s}$
and $W_{m}=-\frac{1}{4\mu_{0}}\int_{s}\vec{B}^{2}\vec{\xi}\cdot\d\vec{s}$
are the virtual works done by the gravitational, surface tension and
magnetic forces, respectively. Since all of these forces, including
the magnetic one in the perfect conductor approximation, are conservative,
the corresponding works can be expressed as the variations of the
associated potential energies. Using the divergence theorem to change
from the surface to volume integrals, and taking into account the
incompressibility constraint $\vec{\nabla}\cdot\vec{\xi}=0$ as well
as the Lagrangian variation $\vec{\xi}\cdot\vec{\nabla}f=\delta f,$
we obtain $W_{g}=-\delta E_{g},$ $W_{s}=-\delta E_{s},$ and $W_{m}=-\delta E_{m},$
where \begin{eqnarray}
E_{g} & = & -\int_{V}\rho\vec{g}\cdot\vec{r}\,\d V,\label{eq:Eh}\\
E_{s} & = & \gamma S,\label{eq:Ec}\\
E_{m} & = & -\frac{1}{4\mu_{0}}\int_{\bar{V}}\vec{B}^{2}\d V\label{eq:Em}\end{eqnarray}
are the associated potential energies. The minus sign at the last
integral is due to the integration over the outer volume $\bar{V}.$
For an equilibrium shape, (\ref{eq:wblnc}) implies $\delta E=0$,
where \begin{equation}
E=E_{g}+E_{s}+E_{m}\label{eq:enrg}\end{equation}
is the total associated energy. This derivation of the energy variation
principle appears more straightforward than the original one by \citet{SneMof82}.
Equilibrium shape of the drop corresponds to a stationary point of
$E,$ which, as usual, has to be a minimum for the equilibrium to
be stable (\citealt{Chan61}).

\subsection{Magnetic energy in a homogeneous external field}

Further, the external magnetic field $\vec{B}_{e}$ is assumed homogeneous,
which supposes the drop to be small compared to the inductor generating
the field. This allows us to express the magnetic energy (\ref{eq:Em})
by an integral over the drop surface as follows. The total magnetic
field is a superposition of the external and induced fields $\vec{B}=\vec{B}_{e}+\vec{B}_{i}.$
Outside the drop, we have $\vec{B}_{i}=-\mu_{0}\vec{\nabla}\Psi_{i},$
where $\Psi_{i}$ is the scalar potential of the induced magnetic
field. Then (\ref{eq:Em}) can be represented as\begin{equation}
E_{m}=E_{0}+E_{1},\label{eq:Em01}\end{equation}
 where \begin{equation}
E_{1}=-\frac{1}{4\mu_{0}}\int_{\bar{V}}\vec{B}\cdot\vec{B}_{i}\,\d V=\frac{1}{4}\int_{S_{\infty}}\Psi_{i}\vec{B}_{e}\cdot\d\vec{s,}\label{eq:E1}\end{equation}
and $S_{\infty}$ is a remote surface enclosing the drop at $r\rightarrow\infty.$
The part of the integral over the drop surface $S$ vanishes because
of the boundary condition (\ref{eq:Bn}). Since the induced magnetic
field is supposed to fall off at large distances $r\rightarrow\infty$
as the dipole field with $\Psi_{i}\sim1/r^{2},$ the last integral
converges to a non-zero value. The other contribution to the magnetic
energy is \begin{equation}
E_{0}=-\frac{1}{4\mu_{0}}\int_{\bar{V}}\vec{B}\cdot\vec{B}_{e}\,\d V=-\frac{1}{4\mu_{0}}\int_{\bar{V}}\vec{B}_{e}^{2}\,\d V-\frac{1}{4\mu_{0}}\int_{\bar{V}}\vec{B}_{i}\cdot\vec{B}_{e}\,\d V,\label{eq:E0}\end{equation}
where the first integral represents the energy of the external magnetic
field, which is constant, and thus negligible, however, formally it
is infinite. Therefore, retaining only the second term in (\ref{eq:E0}),
we obtain\begin{equation}
E_{0}=\frac{1}{4}\int_{\bar{V}}\vec{\nabla}\cdot(\vec{B}_{e}\Psi_{i})\,\d V=-\frac{1}{4}\int_{S}\Psi_{i}\vec{B}_{e}\cdot\d\vec{s}+E_{1},\label{eq:E01}\end{equation}
where the integral is taken over the drop surface $S$ with the outward
normal direction. Now it remains to evaluate the integral in (\ref{eq:E1}),
which is determined by the dipole component of the induced field\[
\Psi_{i}(\vec{r})=\frac{1}{4\pi}\frac{\vec{m}\cdot\vec{r}}{r^{3}},\]
where $\vec{m}=\frac{1}{2}\int_{S}\vec{r}\times\vec{J}\,\d s$ is
the dipole moment of the drop and $\vec{J}$ is the surface current
density. The latter is related to the magnetic field by Ampere's integral
current law, which applied to a small surface element results in \begin{equation}
\vec{J}=\frac{1}{\mu_{0}}\left.\vec{n}\times\vec{B}\right|_{S}=\left.\vec{\nabla}\Psi\times\vec{n}\right|_{S},\label{eq:J}\end{equation}
where $\Psi=\Psi_{e}+\Psi_{i}$ is the full scalar magnetic potential
including also that of homogeneous external field \begin{equation}
\Psi_{e}=-\frac{1}{\mu_{0}}\vec{r}\cdot\vec{B}_{e}.\label{eq:Psi-B0}\end{equation}
 Substituting these expressions into (\ref{eq:E1}), after some algebra
we obtain\begin{equation}
E_{1}=\frac{1}{12}\vec{B}_{e}\cdot\vec{m}=\frac{1}{12}\int_{S}\Psi\vec{B}_{e}\cdot\d\vec{s}.\label{eq:E11}\end{equation}
The integral above can be represented as \[
E_{1}=\frac{1}{12}\left(\int_{S}\Psi_{i}\vec{B}_{e}\cdot\d\vec{s}-\frac{V\vec{B}_{e}^{2}}{\mu_{0}}\right),\]
where the second term, which is related to the energy of homogeneous
external magnetic inside the drop of fixed volume, is constant and,
thus, negligible again. By the same argument, $\Psi_{i}$ in (\ref{eq:E01})
can be substituted by $\Psi.$ Then the magnetic energy (\ref{eq:Em01}),
apart from a constant contribution of the external field, can be written
in terms of $\Psi$ as \begin{equation}
E_{m}=-\frac{1}{4}\int_{S}\Psi\vec{B}_{e}\cdot\d\vec{s}+2E_{1}=-E_{1},\label{eq:Em1}\end{equation}
 where $E_{1}$ is given by (\ref{eq:E11}).

\subsection{Scalar magnetic potential}

There are two alternatives how to find the magnetic potential. First,
the solenoidality constraint $\vec{\nabla}\cdot\vec{B}=0$ for a free-space
magnetic field $\vec{B}=-\mu_{0}\vec{\nabla}\Psi$ results in\begin{equation}
\vec{\nabla}^{2}\Psi=0,\label{eq:Lapl}\end{equation}
which together with the boundary conditions, \begin{equation}
\left.\partial_{n}\Psi\right|_{S}=0,\quad\mbox{and}\quad\left.\Psi\right|_{r\rightarrow\infty}\rightarrow\Psi_{e}=-\mu_{0}^{-1}\vec{r}\cdot\vec{B}_{e}\label{eq:bndc}\end{equation}
governs $\Psi$ outside the drop. This formulation is used in $\S$
\ref{sec:Anal} for analytical treatment of small amplitude deformations
of a circular disk by using a singular Taylor-series-type expansion
around the basic state. Second, for efficient numerical solution,
instead of (\ref{eq:Lapl}), which has to be solved in the whole space
outside the drop, it is more advantageous to use Biot--Savart law
\begin{equation}
\vec{B}(\vec{r})=\vec{B}_{0}-\frac{\mu_{0}}{4\pi}\int_{s}\frac{\vec{r}-\vec{r}'}{|\vec{r}-\vec{r}'|^{3}}\times\vec{J}(\vec{r}')\,\d^{2}\vec{r}',\label{eq:BS}\end{equation}
where the prime denotes the integration point. Then the boundary condition
(\ref{eq:Bn}) applied to (\ref{eq:BS}) results in the surface integral
equation defining $\Psi$ on $S$\begin{equation}
\frac{\mu_{0}}{4\pi}\int_{s}\frac{\vec{r}-\vec{r}'}{|\vec{r}-\vec{r}'|^{3}}\cdot\vec{\nabla}\Psi(\vec{r}')\,\d^{2}\vec{r}'=-\vec{n}\cdot\vec{B}_{e},\label{eq:inteq}\end{equation}
which has to be solved for a given shape of the drop to obtain the
surface distribution of $\Psi,$ which, in turn, defines the magnetic
energy (\ref{eq:E11}). Then equilibrium shape is found by minimising
the total associated energy (\ref{eq:enrg}).

\begin{figure}
\begin{centering}
\includegraphics[width=0.5\columnwidth]{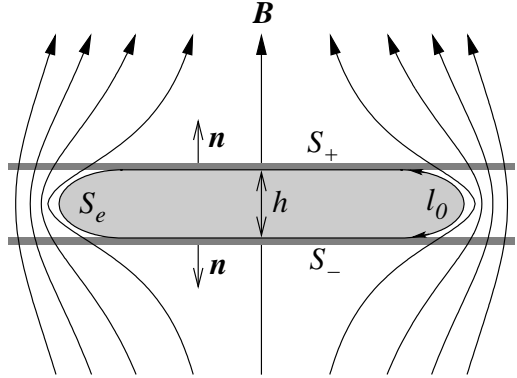} 
\par\end{centering}

\caption{\label{cap:layer}A model of liquid layer confined in the gap between
two horizontal plates in a transverse AC magnetic field.}

\end{figure}

\subsection{Thin-drop model}

In the following, we focus on the case of a thin drop confined in
a horizontal gap between to parallel insulating plates, as shown in
figure \ref{cap:layer}. The drop is modelled by a perfectly conducting
liquid sheet with the virtual displacements constrained to the plane
of the sheet. The external magnetic field $\vec{B}_{e}$ is perpendicular
to the sheet. The surface enclosing the sheet consists of the top
and bottom parts $S_{+}$ and $S_{-},$ both with the same area $S_{0}$
but opposite normals. Taking into account also that the potential
of the induced field changes the sign discontinuously by crossing
the sheet, contributions from both surfaces in (\ref{eq:E11}) are
the same. This results in a factor of two at the front of the integrals
in (\ref{eq:E11}) and (\ref{eq:inteq}) when the integration is carried
out only over the upper part of the sheet. Subsequently, we identify
$S_{0}$ and $\Psi$ with the top surface and the potential at that
surface, respectively. Note that the contribution of the transverse
homogeneous field, whose potential (\ref{eq:Psi-B0}) is constant
along the sheet, vanishes in (\ref{eq:E11}). For the layer of fixed
thickness, gravitational energy is constant and, consequently, irrelevant
in the variation of the total energy. Due to the volume conservation
and fixed thickness, the horizontal area $S_{0}$ is fixed, too. Then
the variation of surface is caused only by the stretching of the perimeter
$P=\oint_{L}\d l,$ which determines the effective edge area $S_{e}=Pl_{0}$
and the corresponding surface energy $E_{s}=\gamma S_{e},$ where
the arclength $l_{0}$ over the edge is assumed to be fixed similar
to the layer thickness itself (see figure \ref{cap:layer}). 

Subsequently, all variables are non-dimensionalised by choosing $R_{0},$
$B_{0},$ $R_{0}B_{0}$ and $\gamma l_{0}R_{0}$ as the length, magnetic
field, potential and energy scales, respectively. Then the dimensionless
associated energy, which comprises a capillary contribution of the
edge and the magnetic energy, can be written as \begin{equation}
E=\oint_{L}\d l-\frac{1}{3}\Bm\int_{S}\Psi\,\d s,\label{eq:nrg-nd}\end{equation}
where $\Bm=B_{0}^{2}R_{0}^{2}/(2\mu_{0}\gamma l_{0})$ is the magnetic
Bond number based on the amplitude of AC magnetic field. Note that
there is no difference between the induced and full magnetic field
potentials in (\ref{eq:nrg-nd}) when that of homogeneous field (\ref{eq:Psi-B0})
is set to be zero along the sheet by a proper choice of additive constant.
Actually, this difference is irrelevant because the contribution of
homogeneous field in (\ref{eq:Psi-B0}), as discussed in the previous
section, is constant for incompressible liquid. For a flat sheet,
(\ref{eq:inteq}) takes the following dimensionless form: \begin{equation}
\frac{1}{2\pi}\int_{s}\frac{\vec{r}-\vec{r}'}{|\vec{r}-\vec{r}'|^{3}}\cdot\vec{\nabla}\Psi(\vec{r}')\,\d^{2}\vec{r}'=-1.\label{eq:inteq2}\end{equation}
For no electric current (\ref{eq:J}) to cross the edge $L,$ $\left.\vec{\tau}\times\vec{n}\cdot\vec{J}\right|_{L}=\left.\partial_{\tau}\Psi\right|_{L}=0$
is required, which implies $\left.\Psi\right|_{L}$=const, where $\vec{\tau}$
and $\vec{\tau}\times\vec{n}$ are the tangent and normal vectors
to the edge, respectively, and $\vec{n}$ is the normal vector to
the sheet$.$ As discussed above, we can set \begin{equation}
\left.\Psi\right|_{L}=0,\label{eq:ecnd}\end{equation}
 which ensures a zero potential for the homogeneous external field
in the plane of the sheet.

\section{\label{sec:Anal}Analytical solution for the stability of circular
disk}

Here the approach developed above will be applied to analyse the stability
of a circular liquid disk with the radius $R=1+R_{1}+R_{2}+\ldots,$
where $R_{1}=\hat{R}_{1}\cos(m\phi)$ is a small perturbation with
the amplitude $\hat{R}_{1}$and the azimuthal wavenumber $m,$ and
$R_{2}$ is a higher-order small correction to be determined later
on. The potential is sought as $\Psi=\Psi_{0}+\Psi_{1}+\Psi_{2}+\ldots,$
where \begin{equation}
\Psi_{0}(\eta,\xi)=-\frac{2}{\pi}\eta\left[1+\xi\arctan(\xi)\right],\label{eq:Psi0}\end{equation}
is the potential of circular disk presented in the angular and radial
oblate spheroidal coordinates, $0\le\eta\le1$ and $0\le\xi<\infty$
(\citealt{LKL02}), which are related with the cylindrical coordinates
by \begin{eqnarray*}
r & = & \sqrt{(1-\eta^{2})(1+\xi^{2})},\\
z & = & \eta\xi.\end{eqnarray*}
Note that $\xi=0$ corresponds to the plane of the disk $z=0,$ where
$r=\sqrt{1-\eta^{2}}$ with $\eta=0$ corresponding to the edge of
a circular disk at $r=1.$ The first-order perturbation of the potential
vanishing away from the disk and satisfying the edge condition (\ref{eq:ecnd}),
which takes the form\begin{equation}
\left.\Psi_{1}\right|_{r\rightarrow1}=-R_{1}\left.\frac{\partial\Psi_{0}}{\partial r}\right|_{r\rightarrow1}=-\frac{2}{\pi}\left.R_{1}\eta^{-1}\right|_{\eta\rightarrow0},\label{eq:disc-edge}\end{equation}
can be written as \begin{equation}
\Psi_{1}(\vec{r})=R_{1}\hat{\Psi}_{1}^{m}(\eta,\xi),\label{eq:Psi1}\end{equation}
 where\begin{equation}
\hat{\Psi}_{1}^{m}(\eta,\xi)=-\frac{2}{\pi}\left(\frac{1-\eta^{2}}{1+\xi^{2}}\right)^{m/2}\frac{\eta}{\eta^{2}+\xi^{2}}.\label{eq:Psi1m}\end{equation}
The details of the solution above, which apart from slightly different
notations are similar to those in \citealt{PEF06}, can be found in
Appendix \ref{sec:apx1}. Since the energy variation about the equilibrium
state is expected to be quadratic in $R_{1},$ we need to consider
also the next-order radius perturbation $R_{2},$ which results from
the area conservation $S=\int_{2\pi}\int_{0}^{R}r\,\d r\d\phi=\pi(1+\hat{R}_{1}^{2}/2+2R_{2}+\ldots)$
as \begin{equation}
R_{2}=-\hat{R}_{1}^{2}/4.\label{eq:R2}\end{equation}
 The second-order potential perturbation, for which the edge condition
(\ref{eq:ecnd}) takes the form \[
\left.\Psi_{2}\right|_{r\rightarrow1}=\left.-R_{1}\frac{\partial\Psi_{1}}{\partial r}-R_{2}\frac{\partial\Psi_{0}}{\partial r}-\frac{R_{1}^{2}}{2}\frac{\partial^{2}\Psi_{0}}{\partial r^{2}}\right|_{r\rightarrow1}=\frac{2}{\pi}R_{2}(1+\cos(2m\phi))\left.(m\eta^{-1}+\eta^{-3})\right|_{\eta\rightarrow0},\]
can be written as $\Psi_{2}(\eta,\xi)=R_{2}\left(\hat{\Psi}_{2}^{0}(\eta,\xi)+\hat{\Psi}_{2}^{2m}(\eta,\xi)\cos(2m\phi)\right).$
Subsequently, we will need only the first term of this expression\[
\hat{\Psi}_{2}^{0}(\eta,\xi)=m\hat{\Psi}_{1}^{0}(\eta,\xi)+\hat{\Psi}_{3}^{0}(\eta,\xi),\]
which satisfies (\ref{eq:Lapl-Psi1m}) with $m=0,$ where $\hat{\Psi}_{1}^{0}(\eta,\xi)$
is defined by (\ref{eq:Psi1m}). The second term above is obtained
similarly to the first one by applying $\partial_{z}^{2}$ to (\ref{eq:Psi0}),
as described in the last paragraph of Appendix \ref{sec:apx1}, which
yields \begin{equation}
\hat{\Psi}_{3}^{0}(\eta,\xi)=-\frac{2}{\pi}\frac{\eta\left[\eta^{2}-\xi^{2}(\xi^{2}+3(1-\eta^{2}))\right]}{(\eta^{2}+\xi^{2})^{3}}.\label{eq:Psi03}\end{equation}
At the disk surface, we have \begin{eqnarray}
\Psi_{0}(r) & = & -\frac{2}{\pi}\sqrt{1-r^{2}},\label{eq:Psi0-r}\\
\hat{\Psi}_{1}^{m}(r) & = & -\frac{2}{\pi}\frac{r^{m}}{\sqrt{1-r^{2}}},\label{eq:Psi1m-r}\\
\hat{\Psi}_{2}^{0}(r) & = & -\frac{2}{\pi}\frac{m+(1-r^{2})^{-1}}{\sqrt{1-r^{2}}}.\label{eq:Psi20-r}\end{eqnarray}
It is important to note that (\ref{eq:Psi1m-r}) and (\ref{eq:Psi20-r})
are singular at $r=1,$ which is the edge of the unperturbed disk.
At the same time, the edge condition (\ref{eq:ecnd}) implies the
potential to be regular (zero) at the actual edge of the deformed
disk. This implies that the solution above can be regularised by representing
it in the radial coordinate $\tilde{r}$ stretched with the radius
of the deformed disk. Using the substitution \begin{equation}
r=R\tilde{r}=(1+\tilde{R})\tilde{r},\label{eq:r-tld}\end{equation}
where $\tilde{R}=R_{1}+R_{2}+\ldots$ is the radius perturbation,
and expanding the solution in power series of $\tilde{R}$ up to the
second order in $R_{1},$ we obtain a solution of the same asymptotic
accuracy, which is free of edge singularities \begin{eqnarray*}
\Psi(r,\phi) & = & \Psi(\tilde{r}(1+\tilde{R}),\phi)\approx\Psi(\tilde{r},\phi)+\tilde{R}\tilde{r}\frac{\partial\Psi}{\partial\tilde{r}}+\frac{(\tilde{R}\tilde{r})^{2}}{2}\frac{\partial^{2}\Psi}{\partial\tilde{r}^{2}}+\ldots=\tilde{\Psi}(\tilde{r},\phi),\end{eqnarray*}
where $\tilde{\Psi}(r,\phi)=\Psi_{0}(r)+R_{1}\tilde{\Psi}_{1}^{m}(r)+R_{2}\left(\tilde{\Psi}_{2}^{0}(r)+\tilde{\Psi}_{2}^{2m}(r)\cos(2m\phi)\right)+\ldots$
and 

\begin{eqnarray*}
\tilde{\Psi}_{1}^{m}(r) & = & -\frac{2}{\pi}\frac{r^{m}-r^{2}}{\sqrt{1-r^{2}}},\\
\tilde{\Psi}_{2}^{0}(r) & = & -\frac{2}{\pi}\left[\frac{(m-1)(1-2r^{m})-r^{2}}{\sqrt{1-r^{2}}}-\frac{2(r^{m}-1)}{(1-r^{2})^{3/2}}\right].\end{eqnarray*}
Then the magnetic energy term in (\ref{eq:nrg-nd}) can be evaluated
up the first order in $R_{2}$ as \[
\int_{0}^{2\pi}\int_{0}^{R}\Psi(r,\phi)r\,\d r\,\d\phi=\int_{0}^{2\pi}R^{2}\int_{0}^{1}\tilde{\Psi}(\tilde{r},\phi)\tilde{r}\,\d\tilde{r}\,\d\phi\approx-\bar{E}_{m}-R_{2}\tilde{E}_{m}.\]
where $\bar{E}_{m}$ and $R_{2}\tilde{E}_{m}$ are the magnetic energies
of circular disk and its leading-order perturbation defined by \begin{eqnarray}
\bar{E}_{m} & = & -2\pi\int_{0}^{1}\Psi_{0}(r)r\,\d r=\frac{4}{3},\label{eq:E0m}\\
\tilde{E}_{m} & = & -2\pi\int_{0}^{1}\left[\tilde{\Psi}_{2}^{0}(r)-4\tilde{\Psi}_{1}^{m}(r)\right]r\,\d r=4(m-1).\label{eq:E1m}\end{eqnarray}
The surface energy term in (\ref{eq:nrg-nd}) is evaluated as \[
\oint_{L}\d l\approx2\pi(1-R_{2}(m^{2}-1))=\bar{E}_{s}+R_{2}\tilde{E}_{s}.\]
 Then the total energy variation is \[
\delta E=\frac{1}{2}R_{2}(\tilde{E}_{s}+\tilde{E}_{m})=-R_{2}(m-1)\left(\pi(m+1)-\frac{2}{3}\Bm\right).\]
Note that there is no energy variation for $m=1$ mode, which corresponds
to the shift of the disk as whole. Circular disk is stable with respect
to small perturbation with $m>1$ as long as its energy is at minimum,
i.e. $\delta E>0.$ Since according to (\ref{eq:R2}) $R_{2}<0,$
the stability condition for $m=2,\,3,\,\ldots.$ is satisfied as long
as \begin{equation}
\Bm\leq3(m+1)\pi/2.\label{eq:Bmc}\end{equation}
The first unstable mode with $m=2,$ which corresponds to an elliptical
deformation, appears as $\Bm$ exceeds the critical value \begin{equation}
\Bm_{c}=\frac{9}{2}\pi.\label{eq:Bm2c}\end{equation}
This critical value is by a factor of 3 greater than the one found
by our previous linear stability analysis (\citealt{PEF06}). The
cause of this discrepancy is discussed in the conclusion of the paper.

\section{\label{sec:Num}Numerical solution}

\begin{figure}
\begin{centering}
\includegraphics[bb=100bp 50bp 360bp 302bp,clip,width=0.5\textwidth]{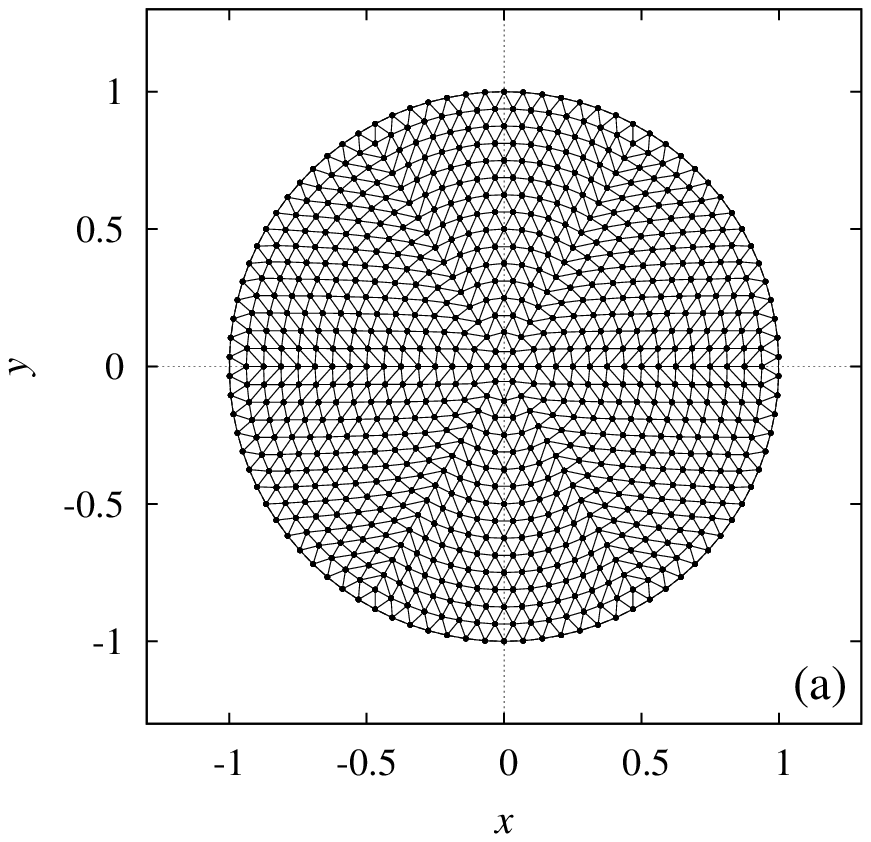}\includegraphics[bb=100bp 50bp 360bp 302bp,clip,width=0.5\textwidth]{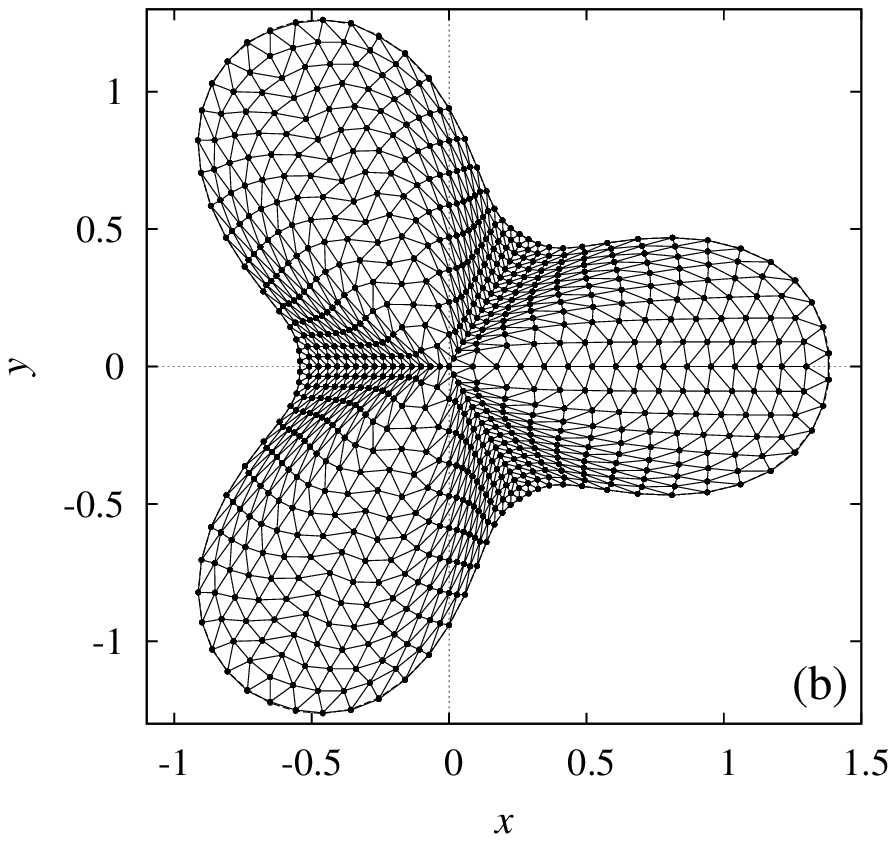}
\par\end{centering}

\caption{\label{fig:mesh}Triangulation of the unit circle with $N=16$ elements
along the radius (a) and a radially stretched mesh fitting the drop
shape (b).}

\end{figure}

This section introduces the numerical method which will be used subsequently
to find equilibrium shapes of thin drops by the approach described
in \S \ref{sec:probform}. Numerical solution will also be verified
against the analytical results obtained in the previous section. To
solve (\ref{eq:inteq2}) with the edge condition (\ref{eq:ecnd}),
which define $\Psi$ over the drop surface $S,$ the latter is tiled
into triangular elements as shown in figure \ref{fig:mesh}(a). Triangulation
is carried out as follows. Firstly, we take a regular hexagon inscribed
in the unit circle and tile it using equilateral triangles with the
side length $1/N.$ Secondly, the hexagon is stretched radially to
fit the unit circle. Then six points are discarded from the perimeter
and the remaining $6(N-1)$ points are redistributed uniformly against
the midpoints of the previous radial level. This produces a more regular
triangulation at the edge, which yields a slightly higher numerical
accuracy. As a result, we obtain a triangular mesh with $6N^{2}-6$
elements and $3N\times(N+1)-5$ vertices. Following the finite element
approach, $\Psi$ is sought at the vertices and interpolated linearly
within the elements. To determine $\Psi$ at the vertices, we need
a corresponding number of equations, which are obtained by numerically
approximating (\ref{eq:inteq2}) at the inner points and applying
the edge condition (\ref{eq:ecnd}) at the peripheral points. The
integral in (\ref{eq:inteq2}) is represented as a sum of integrals
over separate elements, which are approximated by the Gaussian quadratures
for triangles. Thus, for a given mesh $\vec{r}_{i}=(x_{i},y_{i}),$
we obtain a system of linear equations with a dense matrix for unknown
$\Psi_{i}=\Psi(\vec{r}_{i}),$ which are found by the $LU$ decomposition
method.

\begin{figure}
\begin{centering}
\includegraphics[width=0.75\textwidth]{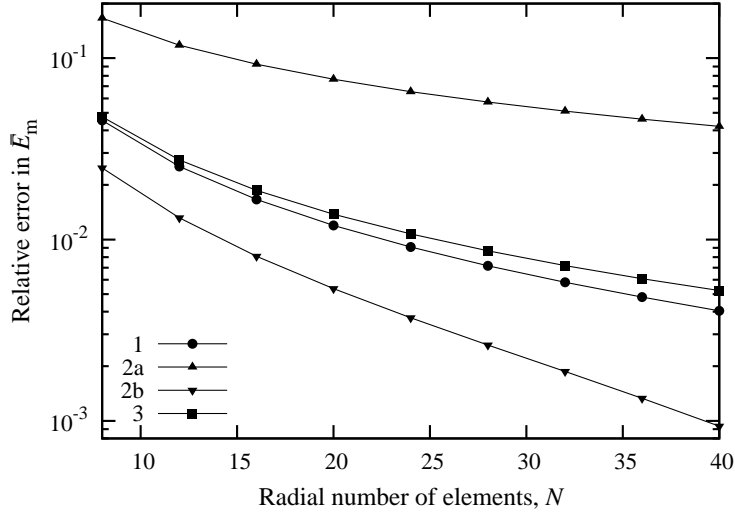} 
\par\end{centering}

\caption{\label{fig:err}The relative error in the magnetic energy of circular
disk (\ref{eq:E0m}) against the radial number of elements $N$ for
four different Gaussian quadratures: (1) linear quadrature using only
the centre point with the weight factor $1;$ (2a) and (2b) quadratic
quadratures using three symmetric points with the barycentric coordinates
$(2/3,1/6,1/6)$ and $(1/2,1/2,0)$, respectively, and the weight
factors $1/3;$ (3) four-point cubic quadrature using the centre point
with the weight factor $-27/48$ and three symmetric points with the
barycentric coordinates $(3/5,1/5,1/5)$ and weight factors $25/48$
(\citealt{Cow73}). }

\end{figure}

The convergence of the magnetic energy for circular disk, $\bar{E}_{m}$
defined by (\ref{eq:E0m}), is shown in figure \ref{fig:err} against
the radial number of elements $N$ for four different quadratures.
Accuracy is lower for the Gaussian quadratures with the evaluation
points located closer to the mesh points. This is because of the integrand
singularities encountered when the observation point belongs to the
element over which the integral is evaluated. Subsequently, we use
a quadratic quadrature with three evaluation points located at the
side midpoints of the element (curve 2b in figure \ref{fig:err}),
which provides the highest accuracy. For the linear elements used
here, the integral in (\ref{eq:inteq2}) can, in principle, be evaluated
exactly. However, such an approach is not applicable because the singularities
between adjacent elements do not cancel out when the current distribution
is a piece-wise constant, as in this case, rather than continuous.
Nevertheless, Gaussian quadratures still provide a reasonably accurate
result also in this case.

\begin{figure}
\begin{centering}
\includegraphics[width=0.75\textwidth]{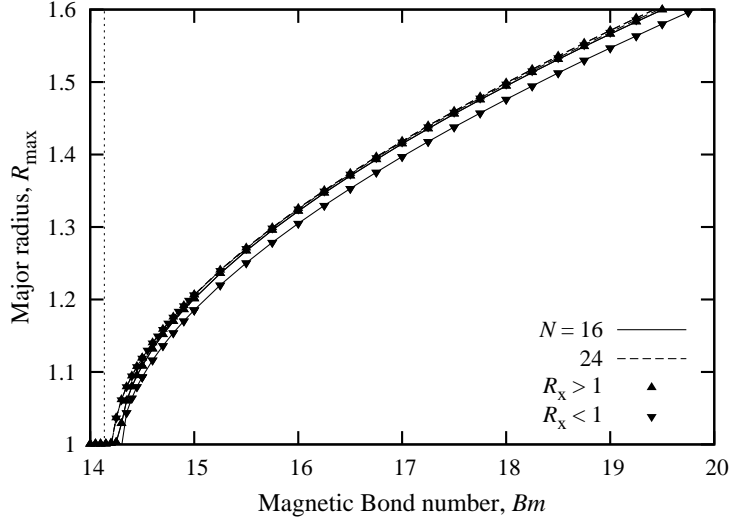} 
\par\end{centering}

\caption{\label{fig:Rmax}Major radius $R_{\max}=\max(R_{x},1/R_{x})$ versus
the magnetic Bond number $\Bm$ for both stretched $(R_{x}>1)$ and
squeezed $(R_{x}<1)$ along the $x$-axis ellipses with $N=16,24$
elements in the radial direction. The vertical dashed line shows the
theoretical threshold value (\ref{eq:Bm2c}). }

\end{figure}

To verify the analytical solution obtained in the previous section,
we first restrict the drop shape to an ellipse defined parametrically
by the mesh point coordinates \begin{equation}
\vec{r}_{i}=(x_{i}^{0}R_{x},y_{i}^{0}/R_{x}),\label{eq:emesh}\end{equation}
where $(x_{i}^{0},y_{i}^{0})=\vec{r}_{i}^{0}$ are the mesh points
for a circular disk and $R_{x}$ is a parameter defining the $x$-radius
of ellipse. For a given $\Bm,$ equilibrium shape is found by using
a Powell-type algorithm (\citealt{NRF96}) to minimise the associated
energy (\ref{eq:nrg-nd}) with respect to $R_{x}.$ For each $R_{x},$
firstly, $\Psi_{i}$ is found by solving the system of linear equations
for the corresponding distribution of mesh points (\ref{eq:emesh}).
Secondly, integrals in (\ref{eq:nrg-nd}) are evaluated numerically
for the given distributions of $\vec{r}_{i}$ and $\Psi_{i}.$

\begin{figure}
\begin{centering}
\includegraphics[width=0.75\textwidth]{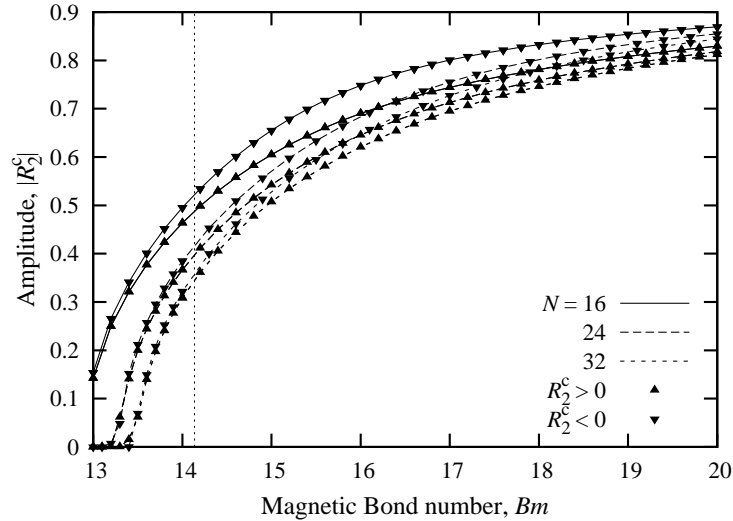} 
\par\end{centering}

\caption{\label{fig:R2-Bm}Amplitude $R_{2}^{c}$ of the radius perturbation
versus the magnetic Bond number $\Bm$ for the drops squeezed along
either the $y$-axis $(R_{2}^{c}>0)$ or $x$-axis $(R_{2}^{c}<0)$
with $N=16,24,32$ elements in the radial direction. The vertical
dashed line shows the theoretical threshold value (\ref{eq:Bm2c}).}

\end{figure}

As seen in figure \ref{fig:Rmax}, which shows the major radius of
ellipse versus $\Bm,$ the critical value of $\Bm,$ by exceeding
which the drop starts to deform, is slightly above its theoretical
value (\ref{eq:Bm2c}). For $N=16$ elements in the radial direction,
the major radius slightly varies depending on whether the ellipse
is stretched $(R_{x}>1)$ or squeezed $(R_{x}<1)$ along the $x$-axis.
Although these two cases differ only by the orientation of the major
axis of ellipse along the $x$- or $y$-axis, which are both theoretically
equivalent, this small difference is due to the six-fold rotational
symmetry of the mesh, which is invariant upon rotation by $60^{\circ}$
but not by $90^{\circ}.$ For $N=24,$ no difference is noticeable
between the $R_{x}>1$ and $R_{x}<1$ cases. 

\begin{figure}
\begin{centering}
\includegraphics[bb=100bp 50bp 360bp 302bp,clip,width=0.5\textwidth]{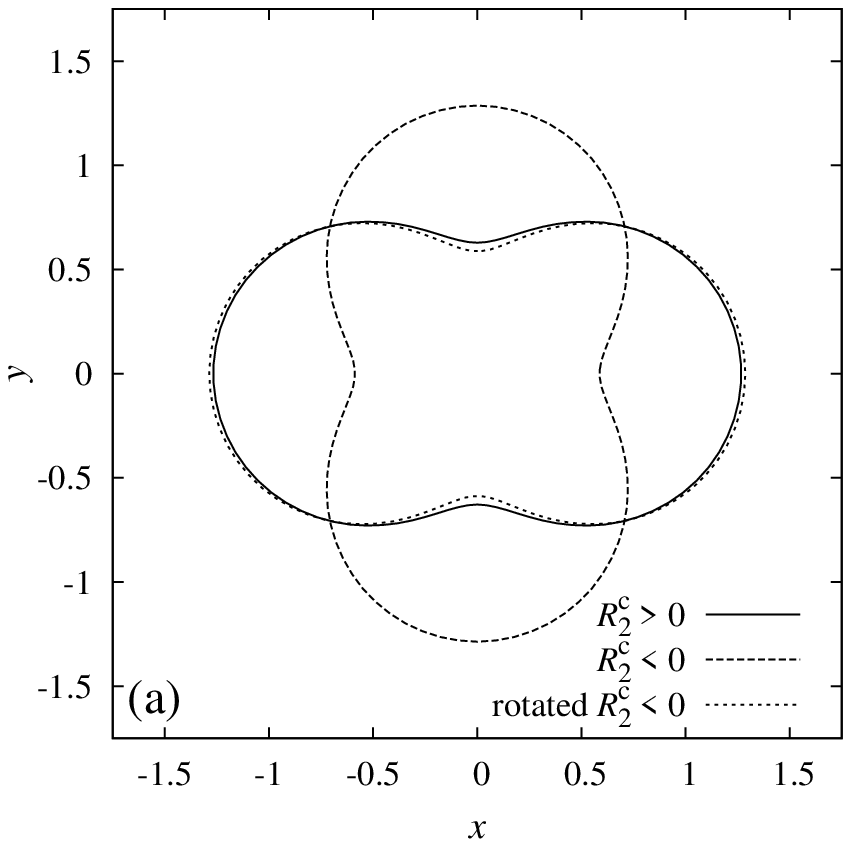}\includegraphics[bb=100bp 50bp 360bp 302bp,clip,width=0.5\textwidth]{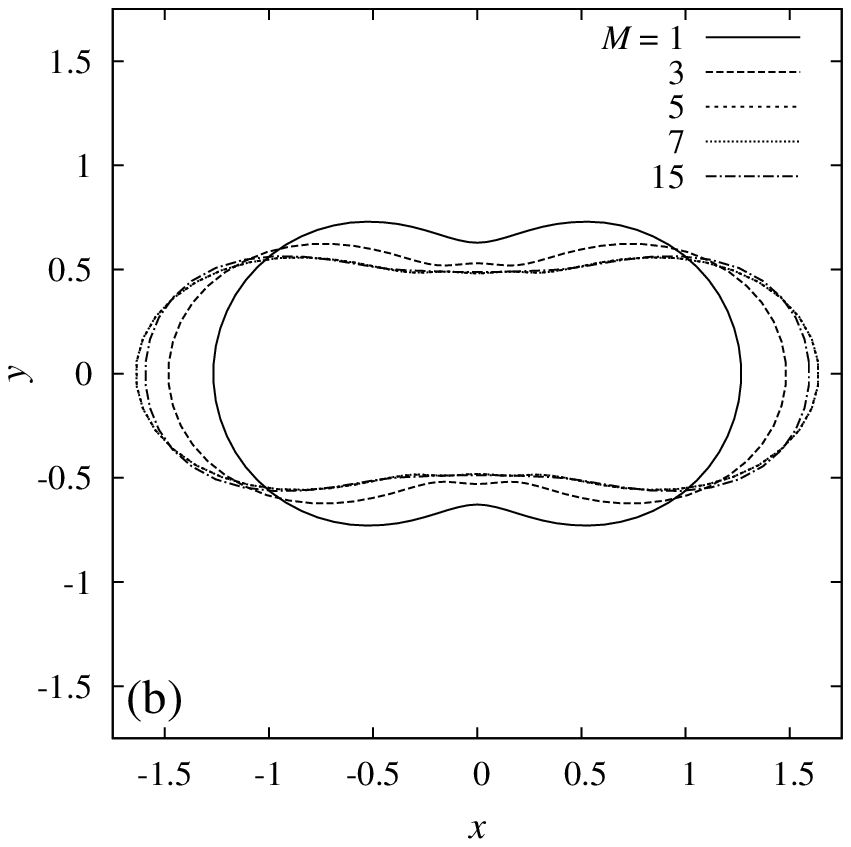}\\
\includegraphics[bb=100bp 50bp 360bp 302bp,clip,width=0.5\textwidth]{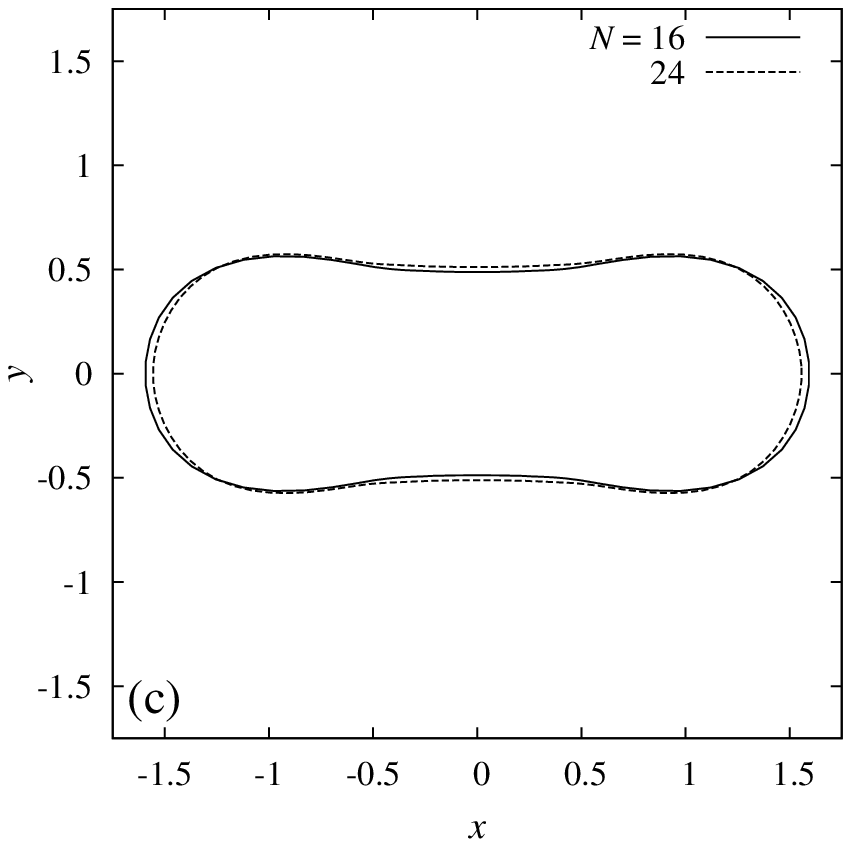} 
\par\end{centering}

\caption{\label{fig:shape-Bm15}Equilibrium shapes at $\Bm=15$ found with
one azimuthal mode $(M=1)$ and $N=16$ elements in radial direction
for the disks squeezed along the $y$- and $x$-axes, which correspond
to $R_{2}^{c}>0$ and $R_{2}^{c}<0,$ respectively (a), $M=1,3,5,7,15$
and $N=16$ for $R_{2}^{c}>0$ (b), $M=15$ with $N=16,24$ for $R_{2}^{c}>0$
(c). }

\end{figure}

Subsequently, we search for the disk radius in the following area-conserving
form\begin{equation}
R^{2}(\phi)=1+\sum_{m=2}^{M+1}[R_{m}^{c}\cos(m\phi)+R_{m}^{s}\sin(m\phi)],\label{eq:R2ser}\end{equation}
where $R_{m}^{c}$ and $R_{m}^{s}$ are unknown amplitudes of cosine
and sine terms in the Fourier series expansion of $R^{2}(\phi).$
Due to the area conservation and the mass centre fixed at the origin,
(\ref{eq:R2ser}) does not contain $m=0$ and $m=1$ terms. Moreover,
owing to the rotational invariance, we can set $R_{2}^{s}=0,$ which
fixes the orientation of the drop up to a rotation by $90^{\circ}$
provided that $R_{2}^{c}\not=0.$ This leaves $2M-1$ unknown coefficients
in (\ref{eq:R2ser}) for the minimisation of the associated energy
(\ref{eq:nrg-nd}). The number of azimuthal modes $M$ is chosen to
ensure the convergence of equilibrium shapes. In this case, the mesh
of unit circle is deformed radially to fit the disk 

\begin{equation}
(r_{i},\phi_{i})=(r_{i}^{0}R(\phi_{i}^{0}),\phi_{i}^{0}),\label{eq:rphi}\end{equation}
where $(r_{i}^{0},\phi_{i}^{0})$ and $(r_{i},\phi_{i})$ are the
polar coordinates of the mesh points for circular and deformed disks,
respectively.

\begin{figure}
\begin{centering}
\includegraphics[bb=100bp 50bp 360bp 302bp,clip,width=0.5\textwidth]{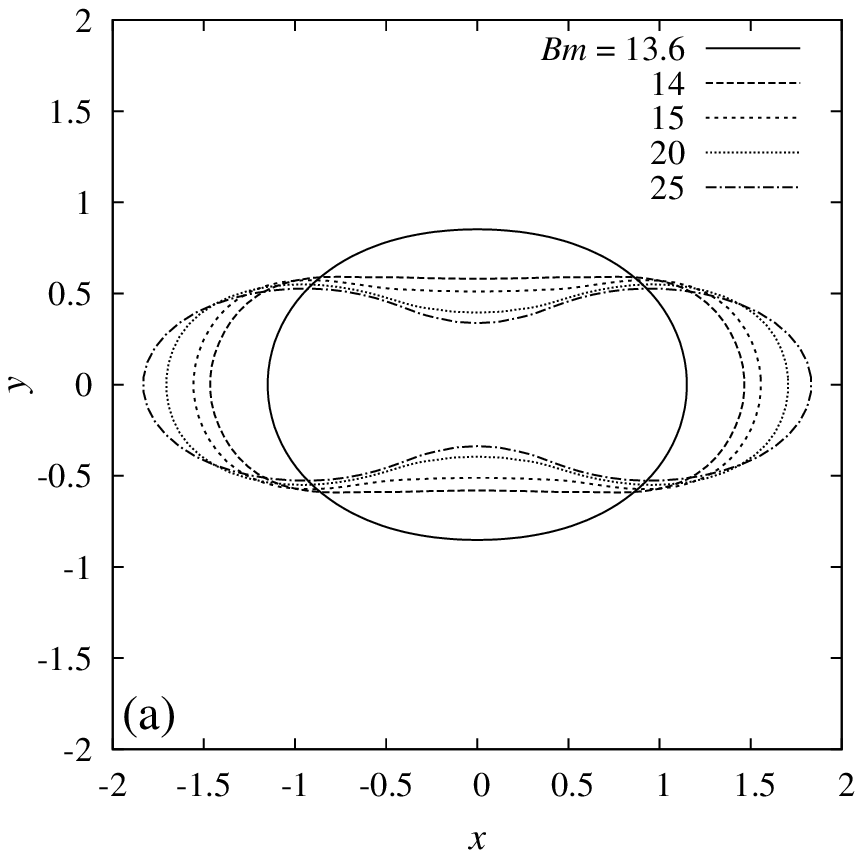}\includegraphics[bb=100bp 50bp 360bp 302bp,clip,width=0.5\textwidth]{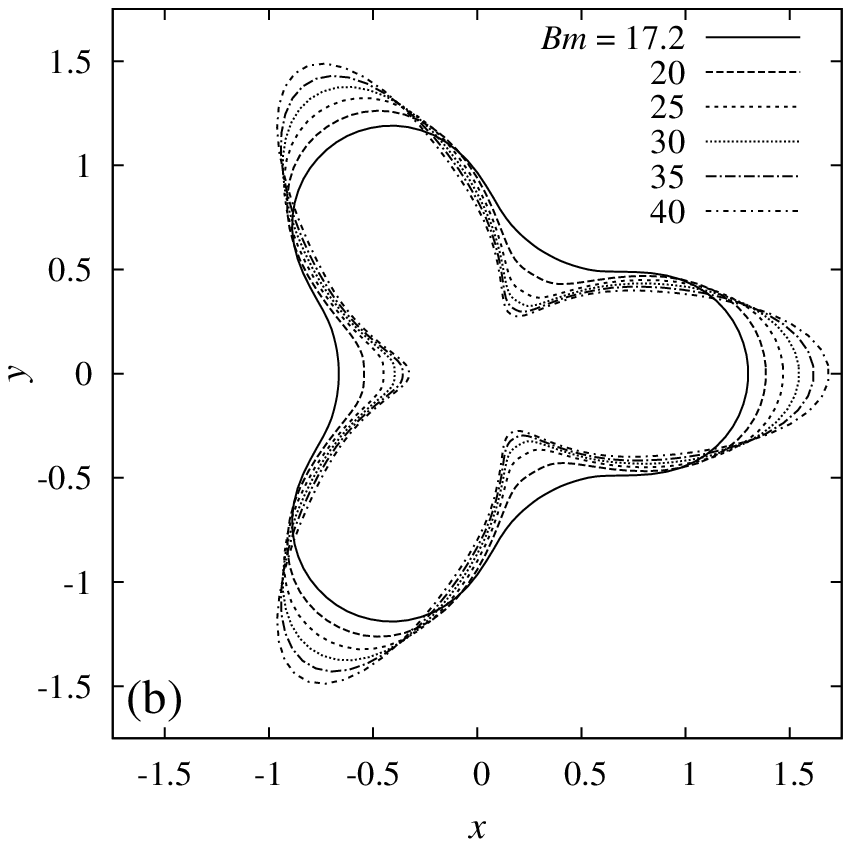} 
\par\end{centering}

\caption{\label{fig:shape-m2-3}Equilibrium shapes for symmetries $m=2$ (a)
and $m=3$ (b) at various magnetic Bond numbers.}

\end{figure}

For comparison with the case of ellipse considered above, we start
with $M=1,$ which leaves only one coefficient, $R_{2}^{c},$ in (\ref{eq:R2ser})
to be determined. As seen in figure \ref{fig:R2-Bm}, which shows
$R_{2}^{c}$ versus $\Bm$ for three numerical resolutions and two
perpendicular orientations of the drop determined by the sign of $R_{2}^{c},$
the radial deformation of the mesh (\ref{eq:rphi}) results in a reduced
numerical accuracy of the critical value of $\Bm,$ which for $N=24$
elements in the radial direction is about $6\%$ lower than its theoretical
value (\ref{eq:Bm2c}). There is also a small difference in the shape
depending on whether the drop is squeezed along the $y$- or $x$-axis
(see figure \ref{fig:shape-Bm15}a). Figures \ref{fig:shape-Bm15}(a)
and (b) show that the shape changes very little as the number of azimuthal
modes and that of the elements in radial direction reach $M=15$ and
$N=24,$ respectively. In the following, we will be using these values
unless stated otherwise.

\begin{figure}
\begin{centering}
\includegraphics[width=0.75\textwidth]{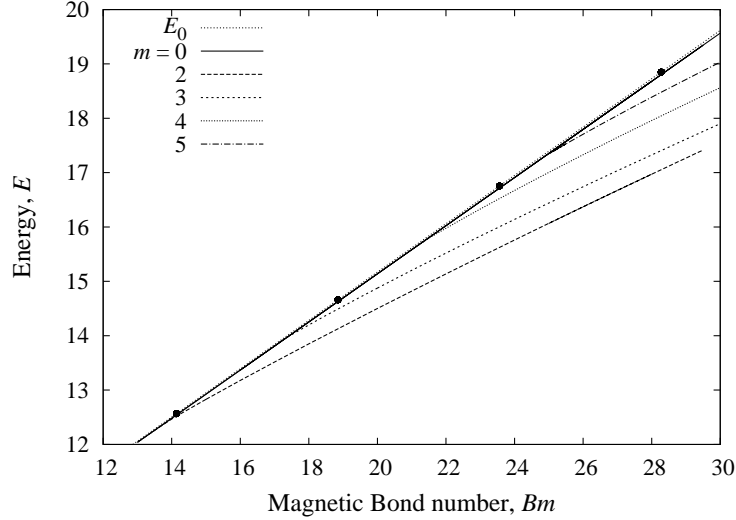} 
\par\end{centering}

\caption{\label{fig:enrgy}The associated energy (\ref{eq:nrg-nd}) versus
$\Bm$ for the shapes of various rotational symmetries. $E_{0}=2\pi+\frac{4}{9}\Bm$
is the associated energy of circular disk. The dots show the analytical
bifurcation points (\ref{eq:Bmc}). }

\end{figure}

The equilibrium shapes found as the magnetic field is gradually increased
are shown in figure \ref{fig:shape-m2-3}a. At $\Bm_{c}\approx13.6,$
which due to the numerical approximation is slightly below the theoretically
predicted stability threshold (\ref{eq:Bm2c}), the drop turns noticeably
elliptic and rapidly elongates with a further increase in $\Bm.$
For $\Bm\gtrsim15,$ the drop starts to tighten around the middle
part. No equilibrium shapes of this type can be found for $\Bm\gtrsim25.$
This implies that the drop may split up into two as the narrowing
of the middle reaches a certain critical value. The splitting of the
drop is not captured by this numerical method, which breaks down as
the neck between two parts of the drop becomes too thin.

\begin{figure}
\begin{centering}
\includegraphics[bb=100bp 50bp 360bp 302bp,clip,width=0.5\textwidth]{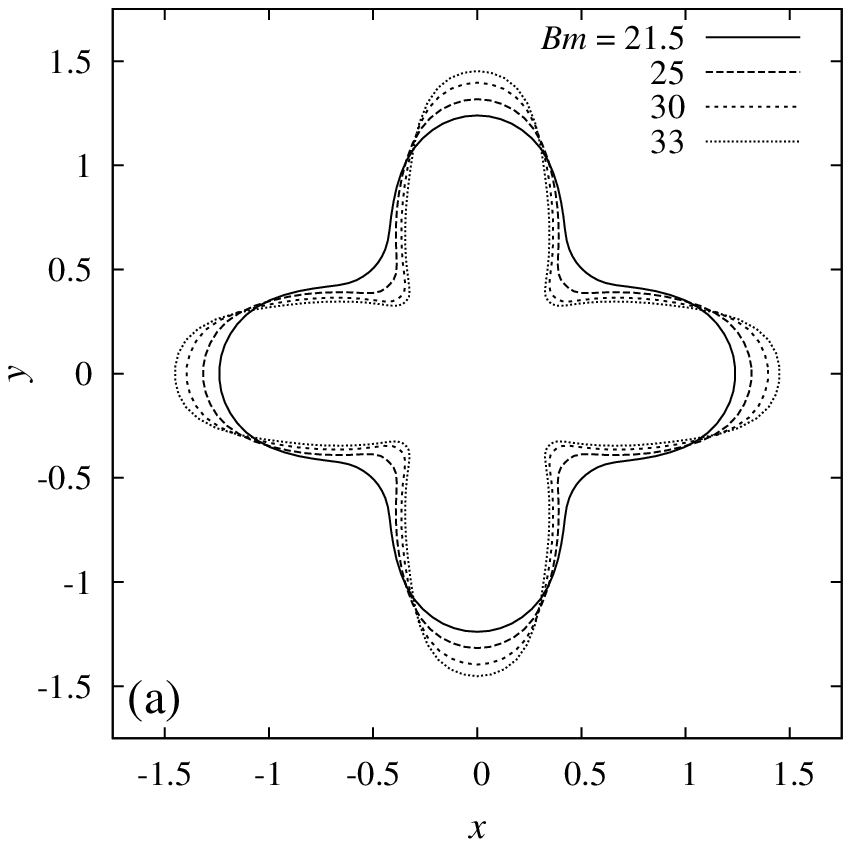}\includegraphics[bb=100bp 50bp 360bp 302bp,clip,width=0.5\textwidth]{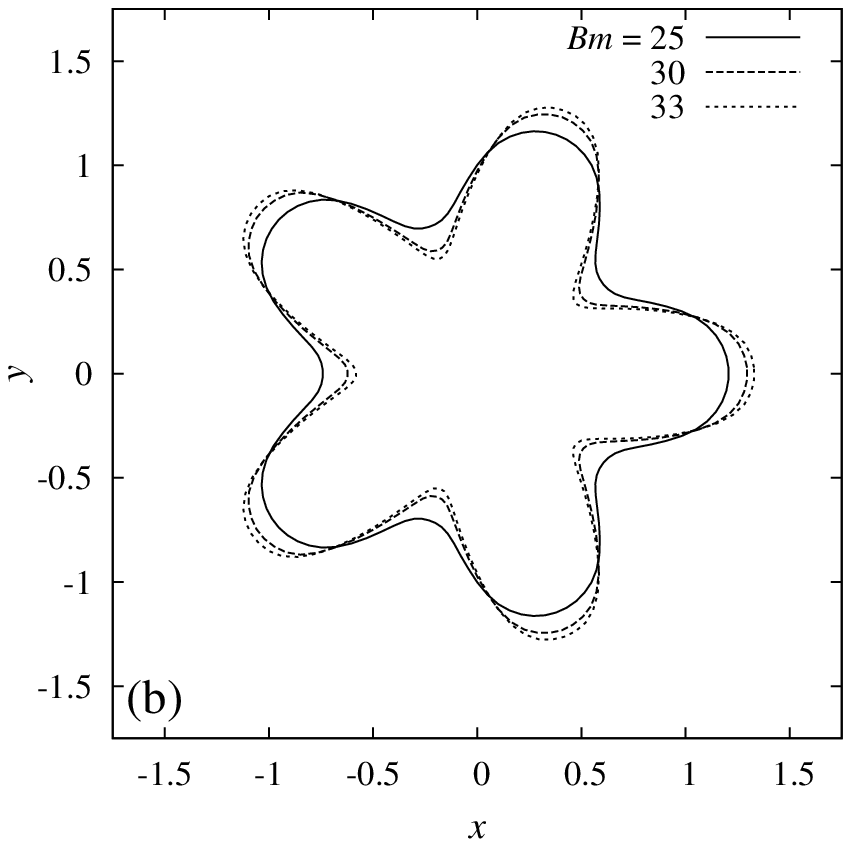} 
\par\end{centering}

\caption{\label{fig:shape-m4-5}Equilibrium shapes for symmetries $m=4$ (a)
and $m=5$ (b) at various magnetic Bond numbers.}

\end{figure}

Alternatively, when the magnetic field is applied instantly with $\Bm\approx20$
to a drop with some initial $m=3$ perturbation, equilibrium shapes
with a three-fold rotational symmetry shown in \ref{fig:shape-m2-3}b
are obtained. As seen in figure \ref{fig:enrgy}, the associated energy
of $m=3$ mode is higher than that of $m=2$ mode, which is also possible
at the same $\Bm.$ Nevertheless, the shapes with three-fold symmetry
are stable because they are separated from the two-fold symmetry shapes
by a finite energy barrier.

\begin{figure}
\begin{centering}
\includegraphics[width=0.5\textwidth]{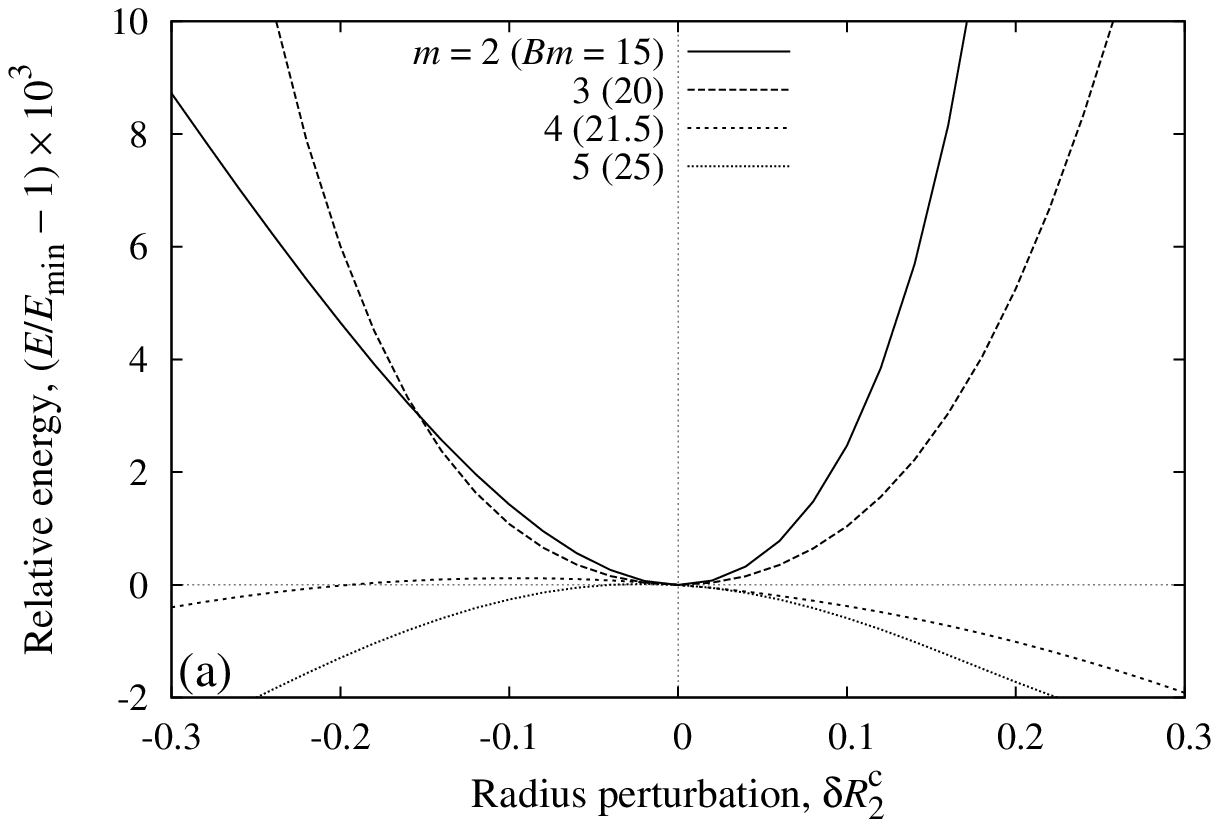}\includegraphics[width=0.5\textwidth]{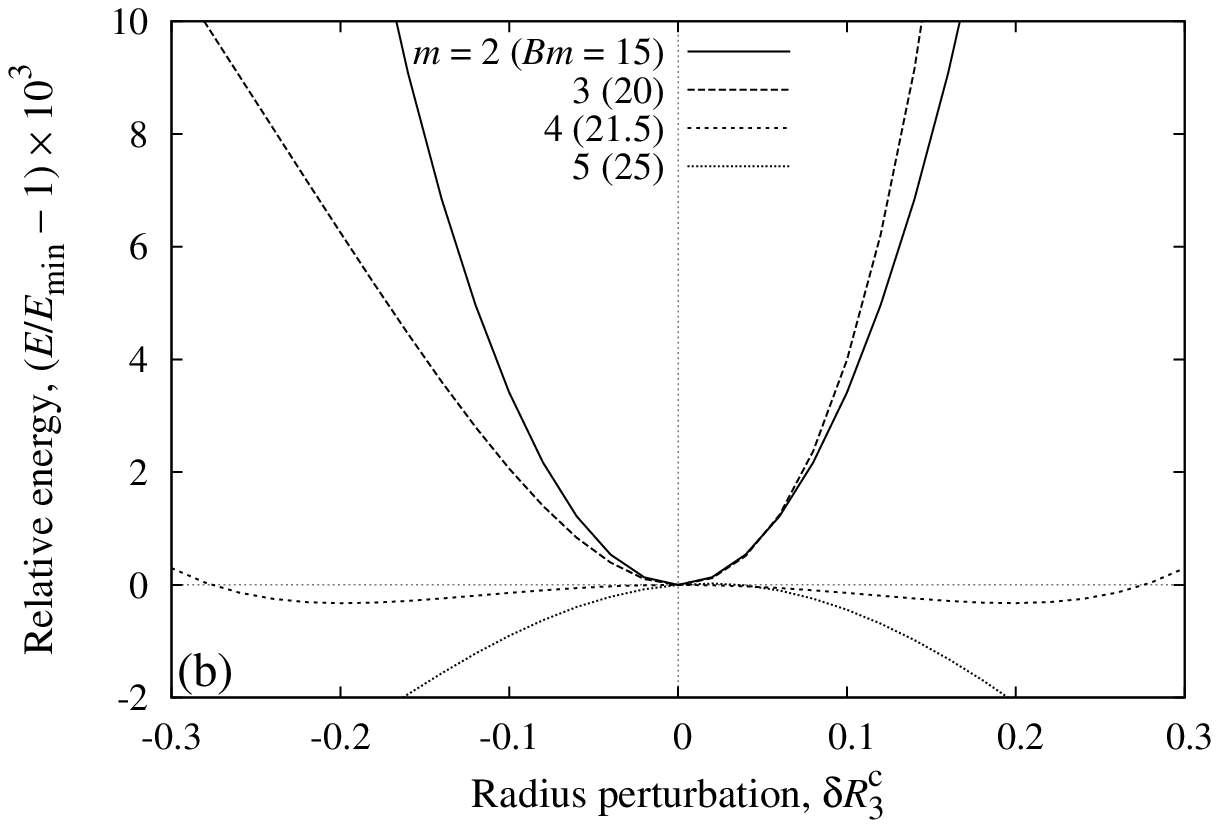} 
\par\end{centering}

\caption{\label{fig:rnrg}The associated energy normalised with its minimum
for near equilibrium shapes of rotational symmetries with $m=2,3,4,5$
versus the perturbation of the cosine amplitude of $m=2$ (a) and
$m=3$ (b) modes defined by $R_{2}^{c}$ and $R_{3}^{c}$ coefficients
in (\ref{eq:R2ser}). }

\end{figure}

This, however, is not the case for the $m=4$ and $m=5$ symmetry
shapes shown in figure \ref{fig:shape-m4-5}(a) and (b), which can
be obtained only when the corresponding symmetry is explicitly imposed
in series (\ref{eq:R2ser}) by ignoring all other modes. As seen in
figure \ref{fig:rnrg}, the associated energy of four- and fivefold
symmetries, in contrast to that of two- and threefold symmetries,
decreases upon $m=2$ and $m=3$ radius perturbations. This implies
that four- and fivefold symmetry shapes are indeed unstable with respect
to these perturbations.

\section{\label{sec:Sum}Summary and conclusions}

In this study, we have numerically modelled strongly deformed equilibrium
shapes of a flat liquid metal drop subject to a transverse high-frequency
AC magnetic field. The drop was treated as a thin liquid layer confined
in a horizontal gap between two parallel insulating plates. AC frequency
was assumed high so that the magnetic field was effectively expelled
from the drop by the skin effect. Equilibrium shapes of the drop were
found by using a variational principle for the associated energy involving
the surface and magnetic contributions. Using Biot--Savart law, the
associated electromagnetic problem was formulated in terms of a surface
integral equation for the scalar magnetic potential. This equation
was solved numerically on an unstructured triangular mesh covering
the surface of the drop. Numerical method was validated against analytical
solution for the stability of circular disk with respect to small-amplitude
azimuthally harmonic edge perturbations. According to the analytical
solution, the edge deformations with the azimuthal wavenumbers $m=2,3,4,\ldots$
start to grow on circular disk as the magnetic Bond number exceeds
the critical threshold values $\Bm_{c}^{(m)}=3\pi(m+1)/2.$ The most
unstable is $m=2$ mode, which corresponds to an elliptical deformation
at the critical Bond number $\Bm_{c}^{(2)}=9\pi/2\approx14.1$. 

This result agrees surprisingly well with the experimental findings
of \citet{CKK06,Con07} for a drop of Galinstan (GaInSn eutectic alloy)
with the diameter of $2R_{0}=65\,\mbox{mm}$ confined in a horizontal
gap between two parallel glass plates separated by $h=3\,\mbox{mm}.$
The drop was submitted to the AC magnetic field generated by a 10-winding
$(n=10)$ coil with the inner and outer radii of $R_{1}=48\,\mbox{mm}$
and $R_{2}=81\,\mbox{mm},$ respectively, which were roughly in the
plane of the drop. For the AC frequency of $f\approx43\,\mbox{kHz},$
which was the highest one applied in the experiment, an originally
circular disk became elliptical as the effective current in the coil
exceeded $I\approx75\,\mbox{A}.$ This corresponds to the r.m.s. magnetic
field in the centre of the coil \[
\frac{B_{0}}{\sqrt{2}}\approx\frac{\mu_{0}nI}{2}\frac{\ln R_{2}-\ln R_{1}}{R_{2}-R_{1}}\approx7.5\,\mbox{mT},\]
which yields the critical Bond number $\Bm=\frac{B_{0}^{2}R_{0}^{2}}{\mu_{0}\pi h\gamma}\approx14,$
where $\gamma=0.718\,\mbox{N/m}$ is the surface tension of Galinstan
and the effective arclength of the edge $l_{0}\approx\pi h/2$ is
approximated by a half circle. The critical currents are higher at
lower frequencies and appear to saturate as the frequency is increased,
which is consistent with the saturation of the electromagnetic force
in the perfect conductor limit. Shapes with a rough three-fold rotational
symmetry are observed above the critical current $I\approx100\,\mbox{A},$
which corresponds to $\Bm\approx25.$ This is by about a third greater
than the theoretical value $\Bm_{c}^{(3)}=6\pi\approx19$ for $m=3$
mode. Note that also the shapes with a four-fold rotational symmetry
are observed in the experiment though the numerical simulation showed
them to be unstable. These discrepancies between the theory and experiment
may be due to two effects. First, the size of the drop is comparable
to that of the coil, which makes the applied magnetic field non-uniform
over the drop radius. Second, to prevent the oxidation the drop is
submerged in a 6\% solution of HCl, which may affect the surface tension.
Given all these experimental uncertainties and deviations from the
idealised theoretical model, the agreement of the instability threshold
for the $m=2$ mode seems too good and perhaps even incidental.

Note that the critical Bond number resulting from the energy variation
approach is by a factor of $3$ greater than that supplied by our
previous linear stability analysis (\citealt{PEF06}). There seem
to be no obvious errors in either approach except for the factor of
$2$ missed in the final expression for the time-averaged force $F_{0}$
above equation (24) of \citet{PEF06}. This factor taken into account
results in $\Bm_{c}^{(m)}=\pi(m+1)/2$ which increases the actual
difference from $1.5$ to $3$ times. The only questionable point
is the determination of electromagnetic force on the edge, where the
magnetic field becomes singular, by the integration of Maxwell stress
tensor over a small cylindrical surface enclosing the edge (\citealt{PEF06}).
It is important to notice that the local magnetic field at the edge
used in the integration is entirely due to the currents induced in
the sheet. Using Ampere's force law, it can be shown that such an
approach accounts only for the interaction between the induced currents
while it misses out any interaction of the induced and external currents.
This is because the latter act via the external magnetic field, which
is opposite to the induced one, but not taken into account by the
local field distribution. As a result, the force on the edge is overestimated
and, consequently, the magnetic field strength necessary for the instability
underestimated. Obviously, the semi-infinite sheet model used by \citet{PEF06}
is not able in principle to account for the interaction with external
magnetic field, which requires the consideration of finite size system.
This is implied also by the energy variation approach, which does
not work for a semi-infinite sheet model. On the one hand, the energy
of the magnetic field, which falls off as $\sim1/\sqrt{r}$ from the
edge, diverges for semi-infinite sheet. On the other hand, this energy
does not vary with the variation of the edge position because this
variation is equivalent to the offset of the origin of coordinate
system. This makes the force on the edge of semi-infinite sheet undetermined.
Such ambiguities do not arise when the energy variation approach is
applied to finite-size drops, as done in this study. Moreover, difficulties
due to the edge singularity disappear altogether when smooth drops
are considered, which, however, significantly increases the numerical
complexity of the problem.

\begin{acknowledgements}

I would like to thank Yves Fautrelle for stimulating discussions.

\end{acknowledgements}

\appendix

\section{\label{sec:apx1}The magnetic potential for harmonically deformed
disk}

In the oblate spheroidal coordinates, (\ref{eq:Lapl}) for the azimuthal
mode $m$ of the potential defined by (\ref{eq:Psi1}) takes the form
\begin{equation}
\frac{\partial}{\partial\eta}\left(\left(1-\eta^{2}\right)\frac{\partial\hat{\Psi}_{1}^{m}}{\partial\eta}\right)+\frac{\partial}{\partial\xi}\left(\left(1+\xi^{2}\right)\frac{\partial\hat{\Psi}_{1}^{m}}{\partial\xi}\right)-\frac{m^{2}\left(\eta^{2}+\xi^{2}\right)}{\left(1-\eta^{2}\right)\left(1+\xi^{2}\right)}\hat{\Psi}_{1}^{m}=0.\label{eq:Lapl-Psi1m}\end{equation}
The potential perturbation, which is supposed to vanish with the distance
from the disk $\left.\hat{\Psi}_{1}^{m}\right|_{\xi\rightarrow\infty}\rightarrow0,$
is related with the radius perturbation by (\ref{eq:disc-edge}),
which now reads as \begin{equation}
\left.\hat{\Psi}_{1}^{m}\right|_{r\rightarrow1}=-\left.\frac{2}{\pi\eta}\right|_{\eta\rightarrow0}.\label{eq:ecnd-m}\end{equation}
Although (\ref{eq:Lapl-Psi1m}) admits the variable separation, such
a solution is complicated by the edge singularity (\ref{eq:ecnd-m}).
Nevertheless, a compact analytical solution can be found similarly
to the construction of spherical solid harmonics from the fundamental
solution of the Laplace equation (\citealt{Batch73}) as follows.
Firstly, note that if $\Psi$ is a solution of the Laplace equation
and $\vec{\epsilon}$ is a constant vector, then $(\vec{\epsilon}\cdot\vec{\nabla})\Psi$
is a solution, too. Secondly, if $\Psi$ satisfies a homogeneous boundary
condition and $\vec{\epsilon}$ is directed along the boundary, then
$(\vec{\epsilon}\cdot\vec{\nabla})\Psi$ satisfies that boundary condition,
too. Thirdly, the operator $(\vec{\epsilon}\cdot\vec{\nabla})$ changes
the radial dependence of $\Psi$ from $\sim(r-1)^{\alpha}$ to $\sim(r-1)^{\alpha-1},$
while the azimuthal dependence is changed from the mode $m$ to $m+1.$
Algebra becomes particularly simple when $\vec{\epsilon}$ is taken
in the complex form as $\vec{\epsilon}=\vec{e}_{x}+\i\vec{e}_{y}=\e^{\i\phi}(\vec{e}_{r}+\i\vec{e}_{\phi}).$
Then each application of $(\vec{\epsilon}\cdot\vec{\nabla})$ is accompanied
by the multiplication with $\e^{\i\phi}.$ Thus, the solution for
$m=1$ is obtained straightforwardly from the axisymmetric base state
(\ref{eq:Psi0}) as \[
\hat{\Psi}_{1}^{1}(\eta,\xi)=-\e^{-\i\phi}\left(\vec{\epsilon}\cdot\vec{\nabla}\right)\Psi_{0}=-\frac{\pi}{2}\left(\frac{1-\eta^{2}}{1+\xi^{2}}\right)^{1/2}\frac{\eta}{\eta^{2}+\xi^{2}}.\]
Higher azimuthal modes can be obtained similarly as $\hat{\Psi}_{1}^{m}=e^{-im\phi}\left(\vec{\epsilon}\cdot\vec{\nabla}\right)^{m}\hat{\Psi}_{0}^{m},$
where $\hat{\Psi}_{0}^{m}$ is an axisymmetric solution satisfying
(\ref{eq:Lapl-Psi1m}). The edge condition (\ref{eq:ecnd-m}) \[
\left(\vec{\epsilon}\cdot\vec{\nabla}\right)^{m}\Psi_{0}^{m}\sim\frac{\Psi_{0}^{m}}{\eta^{2m}}\sim\frac{1}{\eta}\]
yields $\Psi_{0}^{m}\sim\eta^{2m-1}$ for $\eta\rightarrow0.$ Moreover,
the perturbation vanishes far away from the disk when $\left.\hat{\Psi}_{0}^{m}\right|_{\xi=0}=c_{0}^{m}\eta^{2m-1}$
along the whole disk, where $c_{0}^{m}$ is a constant. Then the corresponding
axisymmetric solution of (\ref{eq:Lapl-Psi1m}) can be written as\[
\Psi_{0}^{m}(\eta,\xi)=c_{0}^{m}\sum_{k=1}^{m}c_{k}^{m}P_{2k-1}(\eta)Q_{2k-1}(\i\xi),\]
where $P_{n}(x)$ and $Q_{n}(x)$ are the Legendre polynomials and
functions of the second kind, respectively (\citealt{AbSt72}); the
expansion coefficients are found as $c_{k}^{m}=\frac{4k-1}{Q_{2k-1}(0)}I_{k}^{m},$
where\[
I_{k}^{m}=\int_{0}^{1}\eta^{2m-1}P_{2k-1}(\eta)\,\d\eta=\frac{\sqrt{\pi}2^{1-2m}(2m-1)!}{(m-k)!\Gamma(m+k+1/2)}.\]
 Then the solution for the perturbation amplitude can be written as
\begin{equation}
\hat{\Psi}_{1}^{m}=D_{m-1}^{+}D_{m-2}^{+}\cdots D_{1}^{+}D_{0}\Psi_{0}^{m},\label{eq:Psi1m-dsk}\end{equation}
using the operator\[
D_{m}^{\pm}\equiv\frac{r}{\eta^{2}+\xi^{2}}\left(\xi\frac{\partial\,}{\partial\xi}-\eta\frac{\partial\,}{\partial\eta}\right)\pm\frac{m}{r},\]
which is defined by $D_{m}^{\pm}\equiv\e^{-\i(m\pm1)\phi}\left(\vec{\epsilon}_{\pm}\cdot\vec{\nabla}\right)\e^{\i m\phi},$
where $\vec{\epsilon}_{+}=\vec{\epsilon},$ and $\vec{\epsilon}_{-}=\vec{\epsilon}^{*}$
is the complex conjugate of $\vec{\epsilon}.$ The calculation of
(\ref{eq:Psi1m-dsk}) is algebraically complicated but can be done
by the computer algebra system Mathematica (\citealt{Wolf96}), which
requires considerable computer resources and practically can be carried
out only for $m\leq5.$ But this suffices to deduce the general solution
(\ref{eq:Psi1m}). 

The axisymmetric solution (\ref{eq:Psi03}) with $\sim\eta^{-3}$
edge singularity can be obtained in a similar way directly from the
axisymmetric base solution (\ref{eq:Psi0}) by applying $\left(\vec{\epsilon}_{-}\cdot\vec{\nabla}\right)\left(\vec{\epsilon}_{+}\cdot\vec{\nabla}\right)\equiv D_{1}^{-}D_{0}^{+}.$
This operator is equivalent to $-\partial_{z}^{2}$ because it represents
the transversal part of the Laplace operator while (\ref{eq:Psi0})
satisfies the Laplace equation.

\end{document}